\documentclass[amsmath,amsfonts,twocolumn,showpacs,prb]{revtex4}
\usepackage{graphicx}
\usepackage{epsfig}
\usepackage{bbm}
\usepackage{bm}
\usepackage{dcolumn}
\usepackage{amsmath}
\usepackage{color}
\usepackage[varg]{txfonts}

\newcommand{\beg}{\begin{equation}}
\newcommand{\en}{\end{equation}}
\newcommand{\bp}{\mathbf p}
\newcommand{\bq}{\mathbf q}
\newcommand{\bk}{\mathbf k}
\newcommand{\br}{\mathbf r}

\newcommand{\bn}{\mathbf n}

\newcommand \bel  {\begin{align}}
\newcommand \enl  {\end{align}}

\newcommand{\veps}{\varepsilon}
\newcommand{\eps}{\epsilon}

\newcommand{\up}{\uparrow}
\newcommand{\dn}{\downarrow}
\newcommand{\dg}{^\dagger}

\def\XXint#1#2#3{{\setbox0=\hbox{$#1{#2#3}{\int}$}
     \vcenter{\hbox{$#2#3$}}\kern-.5\wd0}}

\begin{document}

\title{Dynamics of the Schmid-Higgs Mode in $d$-wave superconductors}

\author{Samuel Awelewa and Maxim Dzero}
\affiliation{Department of Physics, Kent State University, Kent, OH 44242, USA}


\date{\today}

\begin{abstract}
We study the dynamics of the longitudinal collective mode in an unconventional superconductor. For concreteness, we assume that the superconductor is described by a $d$-wave order parameter with $d_{x^2-y^2}$ symmetry. After the superconductor has been suddenly subjected to a perturbation at time $t=0$, the order parameter exhibits a peculiar oscillatory behavior, with the amplitude of the oscillations slowly decaying with time in a power-law fashion. Assuming that the initial perturbation is weak, we use a formalism based on quasi-classical approach to superconductivity to determine both the frequency of the oscillations as well as how fast these oscillations decay with time by evaluating the time dependence of the pairing susceptibility. We find that the frequency of the oscillations is given by twice the value of the pairing amplitude in the anti-nodal direction and its amplitude decays as $1/t^2$. The results are also verified by a direct calculation of the order parameter dynamics by numerically solving the equations of motion for the Anderson pseudospins. 
\end{abstract}

\maketitle

\section{Introduction}  
Evolution of the ideas which lead to understanding of collective response of superfluids and superconductors has quite a long history which dates back to 1950s.\cite{GL} These ideas were developed upon a specific realization that the order parameter is a complex function described by an amplitude and a phase.\cite{GL,BCS1957,Anderson1958,Varma2014} While the experimental investigations of the collective excitations associated with the phase fluctuations have been well explored  \cite{Anderson1958,CarlsonGoldman73,Kulik1981} the same could not be said about the collective response associated with the fluctuations of the order parameter amplitude \cite{ASchmid,Galaiko1972,VolkovKogan1973,Kulik1981,Galperin1981,VarmaLit1,VarmaLit2} due to persistent experimental challenges which existed at a time.
In the context of conventional $s$-wave superconductivity these experimental challenges have been recently overcome which not only lead to a series of remarkable experimental results \cite{Pioneers2012,Shimano2013,Shimano2014,THz3,Sherman2015-Disorder,KatoKatsumi2024,Measson2014,Behrle2018,Measson2019,Shimano2020,NonReciprocal2020} but also renewed theoretical interest to this problem.\cite{Papenkort2007,Axt2009,Assa2011,Sachdev2012,Assa2013,Dupuis2014,Manske2014,Cea2016,Moore2017,Silaev2019-Disorder,Joerg2018,Basov2020,Eremin2024,Lorenzana2023,Phan2023,Li2025,Pasha2025} 

Recently there were several experimental reports which study various aspects of the the collective response in unconventional superconductors with $d$-wave order parameter.\cite{KotaHighTc1,KotaHighTc2,Cuprates2020}  Theoretically, a problem of longitudinal collective response in $d$-wave superfluids and superconductors have been addressed by several authors. \cite{BarlasVarma2013,PRL2015,Kirmani2019,Wu2020} While in Refs. \cite{BarlasVarma2013,Wu2020} the main focus has been on computing the value of the Higgs mode frequency,  in Refs.\cite{PRL2015,Kirmani2019} the focus was primarily on the dynamical aspects of the problem. Specifically, the authors of Ref. \cite{PRL2015} have directly analyzed the dynamics of the amplitude mode by solving equations of motion for the normal and anomalous averages following a sudden change of the pairing strength $\lambda\to\lambda'$. In particular, they were able to reproduce the so-called steady state diagram which shows the value of the order parameter at times longer than the order parameter relaxation time $\sim\hbar/\Delta$ as a function of the relative magnitude of the quench $|\lambda'-\lambda|/\lambda$. The resulting steady state diagram turned out to be quite similar to the one found by studying the same problem in the $s$-wave case.\cite{Spivak2004,Levitov2006,Levitov2007} What makes the results of Ref. \cite{PRL2015} interesting is of course the fact that in the $s$-wave case (as well as in the case with $d_{x^2-y^2}+id_{xy}$ pairing) the problem can be solved exactly \cite{Enolski2005,Enolski2005a,Yuzbashyan2006,Yuzbashyan2008,Yuzbashyan2015,Kirmani2019} while for $d_{x^2-y^2}$ pairing it is not exactly solvable and yet in the collisionless regime the superfluid dynamics for both cases appears to be very similar to each other. For example, for quenches above certain critical value order parameter amplitude periodically oscillates with time in both cases.

Despite above mentioned similarities, there are of course important differences such as how fast order parameter asymptotes to a constant value for quenches of small amplitude. For the $s$-wave case it has been well known that the amplitude of the order parameter varies at long times as $\sim\cos(2\Delta_0 t+\pi/4)/\sqrt{t\Delta_0}$, where $\Delta_0$ is the pairing amplitude in equilibrium.\cite{VolkovKogan1973,Enolski2005,Yuzbashyan2015}
At the same time in the case of $d_{x^2-y^2}$ pairing (assuming again the regime of small disturbances) it was only noted that order parameter asymptotes to a constant much faster than in the $s$-wave case without providing more specific details.\cite{PRL2015} The same can be said about the frequency of the collective mode oscillations, i.e. it was not specified whether the order parameter will oscillate with the frequency of $2\Delta_0$ (just like in the $s$-wave case) or some other frequency. This latter question is nontrivial since the order parameter in the $d_{x^2-y^2}$ case has nodes and it is \emph{a priori} not clear what would be the energy of the Schmid-Higgs mode in this case.\cite{BarlasVarma2013}

The goal of this paper is to fill this gap. In the first part of this work we use quasiclassical approach to derive an expression for the longitudinal (Schmid-Higgs) pair susceptibility at zero momentum. By performing the Fourier transform and taking the order parameter in the form $\Delta_{\bn}=\sqrt{2}(n_x^2-n_y^2)\Delta_0$ (here $\bn$ is a unit vector in momentum space) we find that Schmid-Higgs susceptibility at long times oscillates with frequency $2\sqrt{2}\Delta_0$ which implies that the frequency is determined by the pairing gap in the anti-nodal direction. We also find that the amplitude of oscillations decays as $\sim1/(t\Delta_0)^\alpha$ with $\alpha\approx 2$. In the second part of this work we essentially reproduce the results previously reported in Ref. \cite{PRL2015} to confirm our results which we obtained within the quasiclassical approach. 
\section{Formalism}
In what follows we will study the longitudinal collective mode dynamics using the Hamiltonian which assumes that the leading pairing instability is in the $d$-wave channel and neglects the 'off-diagonal in momentum' pairing terms:\cite{PRL2015}
\beg\label{Eq1}
\hat{\cal H}=\sum\limits_{\bp\sigma}(\varepsilon_\bp-\mu)\hat{c}_{\bp\sigma}\dg\hat{c}_{\bp\sigma}-\lambda\sum\limits_{\bp\bq}\gamma_\bp\gamma_\bq
\hat{c}_{\bp\up}\dg\hat{c}_{-\bp\dn}\dg\hat{c}_{-\bq\dn}\hat{c}_{\bq\up}.
\en
Here $\hat{c}_{\bp\sigma}\dg$ ($\hat{c}_{\bp\sigma}$) are fermionic creation (annihilation) operators, $\bp$ is a momentum, $\sigma$ is the spin projection on $z$-axis, $\gamma_\bn=\sqrt{2}({n}_x^2-{n}_y^2)$ is the $d$-wave form factor, which staisfies the normalization condition
\beg\label{NormGamma}
\frac{1}{2\pi}\int\limits_0^{2\pi}|\gamma_\bn|^2{d\phi_\bn}=1, 
\en
$\bn=\bp/p$, $\veps_\bp$ is single particle energy, $\mu$ is a chemical potential and $\lambda$ is the pairing strength. Within the mean-field approximation the superconducting order parameter is given by $\Delta_\bp=\gamma_\bp \Delta$ with 
\beg\label{Delta}
\Delta=\lambda\sum\limits_{\bp}\gamma_\bp\langle c_{-\bp\dn} c_{\bp\up}\rangle.
\en
Within the mean-field approximation the single-particle spectrum is given by $E_\bk=\sqrt{(\veps_\bk-\mu)^2+|\Delta_\bk|^2}$.

Before we proceed, we would like to mention that if one decides to focus explicitly on the study of the collisionless dynamics in the conventional (i.e. fully gapped) superconductors (and fully neglect the possibility that order parameter may develop spacial inhomogeneities) the approximation when one neglects the 'off-diagonal in momentum' pairing terms can actually be justified by employing the Fermi Golden Rule. Specifically, one usually argues that the off-diagonal in momentum terms become effective on a time-scale $\tau_{\mathrm{qp}}\sim\hbar\veps_F/\Delta^2$ which is much longer than the Cooper pair relaxation time $\tau_\Delta\sim\hbar/\Delta$ (here $\veps_F$ is the Fermi energy and $\Delta\ll\veps_F$ is the $s$-wave order parameter). However, since the order parameter for the $d$-superconductor possesses nodes, it is not \emph{a priori} clear if this approximation is justified here also given the existence of the nodal quasiparticles. Specifically, one would expect that due to the excitation of the nodal quasiparticles, the longitudinal (Schmid-Higgs) mode will decay exponentially fast with time. In this regard this situation is somewhat similar to the case of an $s$-wave superconductor with the pair breaking.\cite{Yantao2024,Kamenev2025} The pair breaking leads to the emergence of a single-particle energy scale $\varepsilon_g<\Delta$ and as a result the energy of the Schmid-Higgs mode lies inside the single-particle continuum. Nevertheless in that case the decay of the Schmid-Higgs mode remains power-law due to the square-root singularity in the single-particle density of states. We believe that in this work this approximation is justified for we will focus on studying the dynamics of the Schmid-Higgs mode by evaluating the pair susceptibility at zero momentum. It is however possible that at finite momenta the Schmid-Higgs mode would decay faster just like it happens in the $s$-wave case. \cite{Phan2023,Pasha2025,Kamenev2025} 

\subsection{Eilenberger equation}
In order to describe non-linear response of a $d$-wave superconductor, we are going to use the well established formalism which is based on a quasiclassical approach to superconductivity.\cite{Larkin1965,Eilenberger1968,BELZIG1999,Anton2005} At the foundation of that approach is the Eilenberger equation for the quasiclassical propagator $\check{g}(\bn\eps;\br,t)$: 
\beg\label{Eilenberger}
\begin{aligned}
&[\veps\check{\tau}_3-\check\Delta_\bn(\br,t)\stackrel{\circ},\check{g}]+\frac{i}{2}\left\{\check{\tau}_3,\partial_t\check{g}\right\}+\frac{ev_F}{c}[\bn{\mathbf A}(\br,t)\check{\tau}_3\stackrel{\circ},\check{g}]\\&=-i{v}_F(\bn\mbox{\boldmath $\nabla$}_\br)\check{g},
\end{aligned}
\en
where 
\beg\label{vectorA}
{\mathbf A}(\br,t)=\left(\frac{c{\mathbf E}}{i\omega}\right)e^{i(\bk\br-\omega t)}+\textrm{c.c.}
\en
is an external vector potential and ${\mathbf E}$ is the amplitude of an external electric field. The quasiclassical matrix propagator is a four-by-four matrix defined in the Nambu and Keldysh subspaces:\cite{VolkovKogan1973,Kamenev2009,Kamenev2011} 
\beg\label{KeldyshProps}
\check{g}=\left[\begin{matrix} \hat{g}^R & \hat{g}^K \\ 0 & \hat{g}^A\end{matrix}\right]
\en
and is a subject to normalization condition
\beg\label{norm}
\check{g}\cdot\check{g}=\check{{\mathbbm{1}}}.
\en
The convolution should be understood as follows
\beg\label{SecondTerm}
\begin{aligned}
\left[\check{\Delta}\stackrel{\circ},\check{g}\right]=&\int\frac{d\eps}{2\pi}\left\{\check{\Delta}(\br,t)e^{-\frac{i}{2}\stackrel{\leftarrow}\partial_t\stackrel{\rightarrow}\partial_\eps}\check{g}(\bn\eps;\br,t)\right.\\&\left.-
\check{g}(\bn\eps;\br,t)e^{\frac{i}{2}\stackrel{\leftarrow}\partial_\eps\stackrel{\rightarrow}\partial_t}\check{\Delta}(\br,t)\right\}e^{-i\eps(t_1-t_2)}
\end{aligned}
\en
and $\check{\Delta}$ is diagonal in Keldysh subspace.
In what follows we will work in the clean limit and, as a consequence, equation (\ref{Eilenberger}) does not include the self-energy part which accounts for the disorder effects. 
\subsection{Ground state}
In the ground state
\beg\label{DLTgs}
\hat{\Delta}_\bn(\br,t)=\left(i\hat{\tau}_2\right)\gamma_\bn\Delta\equiv i\hat{\tau}_2\Delta_\bn
\en
and 
\beg\label{ggsRA}
\hat{g}_{\bn\eps}^{R(A)}=\hat{\tau}_3 g_{\bn\eps}^{R(A)}-i\hat{\tau}_2f_{\bn\eps}^{R(A)},
\en
Here $\hat{\tau}_a$ are the Pauli matrices which act in Nambu space. Given the normalization condition (\ref{norm}) the Keldysh component is a simple parametrization
\beg\label{ggsK}
\hat{g}_{\bn\eps}^{K}=\left(\hat{g}_{\bn\eps}^{R}-\hat{g}_{\bn\eps}^{A}\right)\tanh\left(\frac{\eps}{2T}\right)
\en
and $T$ is temperature.
Given the matrix relation $\hat{\tau}_2\hat{\tau}_3-\hat{\tau}_3\hat{\tau}_2=2i\hat{\tau}_1$,
for the retarded components of the propagator we have
\beg\label{etaeps}
g_{\bn\eps}^{R(A)}=\frac{\eps}{\eta_{\bn\eps}^{R(A)}}, \quad 
f_{\bn\eps}^{R(A)}=\frac{\Delta_{\bn}}{\eta_{\bn\eps}^{R(A)}}.
\en
Functions $\eta_{\bn\eps}^{R(A)}$ are given by 
\beg\label{etaRA}
\eta_{\bn\eps}^{R(A)}=\left\{\begin{split} &\pm\mathrm{sign}(\eps)\sqrt{(\eps\pm i\delta)^2-\Delta_{\bn}^2}, \quad |\eps|\geq|\Delta_\bn|, \\
&i\sqrt{\Delta_\bn^2-\eps^2}, \quad |\eps|<|\Delta_\bn|.
\end{split}
\right.
\en
Given the self-consistency equation (\ref{Delta}) the value of the pairing amplitude in equilibrium is determined by the solution of the following equation
\beg\label{GetDLTeq}
\Delta=\frac{\lambda}{2}\int\limits_0^{2\pi}\frac{d\phi_\bn}{2\pi}\gamma_\bn\int\limits_{-\infty}^\infty{d\eps}\textrm{Tr}\left\{-i\hat{\tau}_2\hat{g}_{\bn\eps}^K\right\}.
\en

\subsection{Perturbative solution of the Eilenberger equation}
Our intermediate goal will be to compute the second order corrections to the quasiclassical propagator $\check{g}$ in powers of the vector potential. We keep the vector potential primarily as a book keeping tool. Indeed, as a reader may have already realized, the calculation of the pair susceptibility will require the calculation of the corrections to the pairing field $\check{\Delta}_\bn(\br,t)$. Since the pairing amplitude is a scalar quantity, the corresponding corrections will be determined by the second order corrections to $\check{g}$ in powers of the vector potential. Consequently, we will keep the external vector potential to compute the first order corrections to $\check{g}$. 

\paragraph{First order corrections.} Given \eqref{vectorA} we will look for the first order corrections to the retarded and advanced parts $\hat{g}_1^{R(A)}(\bn\eps;\br t)$ in the form
\beg\label{g1four}
\hat{g}_1^{R(A)}(\bn\eps;\br t)=\hat{g}_1^{R(A)}(\bn\eps;\bk\omega)e^{i(\bk\br-\omega t)}.
\en
Equation which determines the first order correction to the retarded and advanced components of $\hat{g}_1^{R(A)}(\bn\eps;\bk\omega)$ is
\beg\label{Eq4g1}
\begin{aligned}
&[\veps\hat{\tau}_3-\hat\Delta_\bn,\hat{g}_1]+\frac{1}{2}\left\{{\omega}\hat{\tau}_3-{v}_F(\bn\bk)\hat{\tau}_0,\hat{g}_1\right\}\\&=
\left(\frac{ev_F}{i\omega}\right)(\bn{\mathbf E})\left[\hat{g}_{\bn\eps+\omega/2}\hat{\tau}_3-
\hat{\tau}_3\hat{g}_{\bn\eps-\omega/2}\right]
\end{aligned}
\en
(here we omitted the superscripts for brevity).
Note that the first term (\ref{Eq4g1}) is approximate. The reason is that $\hat{\Delta}_\bn={\gamma}_{\bn}\hat{\Delta}$ with $\bn=\bp/p_F$  is actually a function of momentum \eqref{SecondTerm}. Given the Groenewold-Moyal product rule:
\beg\label{GMoyal}
\begin{aligned}
\hat{A}\circ\hat{B}=\hat{A}_{\bp\eps}(\br,t)e^{\frac{i}{2}\left(\stackrel{\leftarrow}\partial_\br\stackrel{\rightarrow}\partial_\bp-\stackrel{\leftarrow}\partial_t\stackrel{\rightarrow}\partial_\eps-\stackrel{\leftarrow}\partial_\bp\stackrel{\rightarrow}\partial_\br+\stackrel{\leftarrow}\partial_\eps\stackrel{\rightarrow}\partial_t\right)}\hat{B}_{\bp\eps}(\br,t), 
\end{aligned}
\en
after the Wigner transformation expression \eqref{SecondTerm} acquires the following form
\beg\label{ApplyGMoyal}
\begin{split}
&\hat{\Delta}_\bp\exp\left[\frac{i}{2}\left(\stackrel{\leftarrow}\partial_\br\stackrel{\rightarrow}\partial_\bp-\stackrel{\leftarrow}\partial_t\stackrel{\rightarrow}\partial_\eps-\stackrel{\leftarrow}\partial_\bp\stackrel{\rightarrow}\partial_\br+\stackrel{\leftarrow}\partial_\eps\stackrel{\rightarrow}\partial_t\right)\right]\hat{g}_{1}(\bp\eps;\br,t)\\&=\hat{\Delta}_\bp\exp\left[-\frac{i}{2}\stackrel{\leftarrow}\partial_\bp\stackrel{\rightarrow}\partial_\br\right]\hat{g}_{1}(\bp\eps;\br,t)=\hat{\Delta}_{\bp+\frac{\bk}{2}}\hat{g}_{1}(\bp\eps;\br,t).
\end{split}
\en
Since $\Delta_\bp$ is evaluated at $\bp=p_F\bn$ in the case when $p_F\gg k$ we can ignore its dependence on $\bk$. Furthermore, function $\hat{g}_1^{R(A)}$ satisfies the normalization condition 
\beg\label{Norm1}
\hat{g}_{\bn\eps+\frac{\omega}{2}}^{R(A)}\hat{g}_{1}^{R(A)}(\bp\eps;\bk,\omega)+\hat{g}_{1}^{R(A)}(\bp\eps;\bk,\omega)\hat{g}_{\bn\eps-\frac{\omega}{2}}^{R(A)}=0.
\en
We will be primarily interested in the solution of equation (\ref{Eq4g1}) for $\bk=0$:
\beg\label{g1RAfin}
\begin{aligned}
\hat{g}_{1}^{R(A)}(\bn\eps;0,\omega)&=\left(\frac{ev_F}{i\omega}\right)\frac{\left[\hat{\tau}_3-
\hat{g}_{\bn\eps+\frac{\omega}{2}}^{R(A)}\hat{\tau}_3\hat{g}_{\bn\eps-\omega/2}^{R(A)}\right](\bn{\mathbf E})}{\left(\eta_{\bn\eps+\frac{\omega}{2}}^{R(A)}+\eta_{\bn\eps-\frac{\omega}{2}}^{R(A)}\right)}.
\end{aligned}
\en
Consequently, it is straightforward to show that the linear correction to the Keldysh component is given by 
\beg\label{g1k}
\hat{g}_1^K=\hat{g}_1^Rt_{\eps-\frac{\omega}{2}}-\hat{g}_1^At_{\eps+\frac{\omega}{2}}
\en
with $t_\eps=\tanh\left({\eps}/{2T}\right)$.
\paragraph{Second order corrections: retarded and advanced components.} We proceed with the calculation of the second order correction to the retarded and advanced components of $\check{g}$. As it has been already mentioned above there is a correction to the longitudinal component of the pairing field which we represent as
\beg\label{deltaDelta}
\delta\hat{\Delta}_\bn(\br,t)=\left(-i\hat{\tau}_2\right)\delta\Delta_\bn^L(\bq,\nu)e^{i(\bq\br-\nu t)}\equiv\delta\hat{\Delta}_{\bn}^L\cdot e^{i(\bq\br-\nu t)}.
\en
Given definition \eqref{vectorA} we obviously have three different cases to consider: (1) $\bq=2\bk$, $\nu=2\omega$; (2) $\bq=0$, $\nu=0$ and (3) $\bq=-2\bk$, $\nu=-2\omega$. As it will become clear below, we will be mainly interested in computing the corrections in the limit $\bk\to0$ and also at the end of the calculation we will take ${\mathbf E}\to 0$. For this reason it will suffice for us to consider the first case:
\beg\label{g2}
\hat{g}_2(\bn\eps;\br t)=\hat{g}_2(\bn\eps;\bk\omega)e^{2i(\bk\br-\omega t)}.
\en
Then equation for the function $\hat{g}_2^{R(A)}(\bn\eps;\bk\omega)$ reads
\beg\label{Eq4g2a}
\begin{split}
&[\veps\hat{\tau}_3-\hat\Delta_\bn,\hat{g}_2]-2{v}_F(\bn\bk)\hat{\tau}_0\hat{g}_2+\omega\left\{\hat{\tau}_3,\hat{g}_2\right\}=\left[\delta\hat{\Delta}_\bn^L\stackrel{\circ},\hat{g}_{0}\right]\\&+
\left(\frac{ev_F}{i\omega}\right)(\bn{\mathbf E})\left[\hat{g}_{1}(\bn\eps+\omega/2;\bk\omega)\hat{\tau}_3-
\hat{\tau}_3\hat{g}_{1}(\bn\eps-\omega/2;\bk\omega)\right].
\end{split}
\en
Given \eqref{deltaDelta} the commutator on the right hand side of this equation is given by
\beg\label{CommLHS}
\left[\delta\hat{\Delta}_\bn^L\stackrel{\circ},\hat{g}_{0}\right]=\delta\hat{\Delta}_\bn^L\hat{g}_{\bn\eps-\omega}-\hat{g}_{\bn\eps+\omega}\delta\hat{\Delta}_\bn^L.
\en
Function $\hat{g}_2^{R(A)}(\bn\eps;\bk\omega)$ satisfies the following normalization condition
\beg\label{normg2RA}
\begin{aligned}
&\hat{g}_{\bn\eps+\omega}\hat{g}_2(\bn\eps;\bk\omega)+\hat{g}_2(\bn\eps;\bk\omega)\hat{g}_{\bn\eps-\omega}\\&+\hat{g}_1(\bn\eps+\omega;\bk\omega)\hat{g}_1(\bn\eps-\omega;\bk\omega)=0.
\end{aligned}
\en
Then solution for the retarded and advanced components of $\check{g}_2$ at $\bk=0$ and in the limit ${\mathbf E}\to 0$ reads
\beg\label{g2RAFinal}
\begin{aligned}
&\hat{g}_2^{R(A)}(\bn\eps;0,\omega)=\frac{\left[\hat{g}_{\bn\eps+\omega}^{R(A)}\delta\hat{\Delta}_\bn^L\hat{g}_{\bn\eps-\omega}^{R(A)}-\delta\hat{\Delta}_\bn^L\right]}{\eta_{\bn\eps+\omega}^{R(A)}+\eta_{\bn\eps-\omega}^{R(A)}}.
\end{aligned}
\en
\paragraph{Second order corrections: Keldysh component.}
Equation for the $\hat{g}_2^K$ is of course exactly the same as \eqref{Eq4g2a}.
However, the solution for $\hat{g}_2^K$ is different from $\hat{g}_2^{R(A)}$ because it satisfies the different normalization condition:
\beg\label{norm4g2K}
\begin{split}
&\hat{g}_{\eps+\omega}^R\hat{g}_2^K+\hat{g}_{\eps+\omega}^K\hat{g}_2^A+\hat{g}_2^K\hat{g}_{\eps-\omega}^A+\hat{g}_2^R\hat{g}_{\eps-\omega}^K\\&+\hat{g}_1^R(\bn\eps+\omega/2;\bk\omega)\hat{g}_1^K(\bn\eps-\omega/2;\bk\omega)\\&+\hat{g}_1^K(\bn\eps+\omega/2;\bk\omega)\hat{g}_1^A(\bn\eps-\omega/2;\bk\omega)=0.
\end{split}
\en
Similar to the ansatz \eqref{g1k} we will look for $\hat{g}_2^K$ in the following form
\beg\label{g2Kansatz}
\begin{aligned}
\hat{g}_2^K(\bn\eps;\bk\omega)&=\hat{g}_2^R(\bn\eps;\bk\omega)t_{\eps-\omega}-t_{\eps+\omega}\hat{g}_2^A(\bn\eps;\bk\omega)\\&+\delta \hat{g}_2^K(\bn\eps;\bk\omega).
\end{aligned}
\en
The main advantage of using \eqref{g2Kansatz} is that the normalization condition \eqref{norm4g2K} significantly simplifies:
\beg\label{norm4dg2K}
\hat{g}_{\eps+\omega}^R\delta\hat{g}_2^K+\delta\hat{g}_2^K\hat{g}_{\eps-\omega}^A=0.
\en
the resulting expression for the function $\delta\hat{g}_2^K$ evaluated at $\bk=0$ and for ${\mathbf E}\to0$ reads
\beg\label{dg2KFinal}
\begin{aligned}
\delta\hat{g}_2^K(\bn\eps;0,\omega)&=\frac{\left(\delta\hat{\Delta}_\bn^L-\hat{g}_{\bn\eps+\omega}^R\delta\hat{\Delta}_\bn^L\hat{g}_{\bn\eps-\omega}^A\right)}{\eta_{\bn\eps+\omega}^{R}+\eta_{\bn\eps-\omega}^{A}}f(\eps,\omega),
\end{aligned}
\en
where $f(\eps,\omega)=t_{\eps-\omega}-t_{\eps+\omega}$.
Expressions (\ref{g2RAFinal},\ref{g2Kansatz}) along with \eqref{dg2KFinal} are the main results of this Section.

\section{Longitudinal pair susceptibility} 
The expression for the Schmid-Higgs (SH) $d$-wave  susceptibility can now be derived by employing the self-consistency equation 
\beg\label{SelfEq}
\delta\Delta_{\omega}^L=\frac{\lambda}{2}\int\limits_0^{2\pi}\frac{d\phi_\bn}{2\pi}\gamma_\bn\int\limits_{-\infty}^\infty{d\eps}\textrm{Tr}\left\{-i\hat{\tau}_2\hat{g}_2^K(\bn\eps;0,\omega)\right\},
\en
In the expression for $\hat{g}_2^K(\bn\eps;0,\omega)$ we will not need the terms which are explicitly proportional to the electric field and this is why we set the external electric field to zero, ${\mathbf E}=0$, in the expressions (\ref{g2RAFinal},\ref{g2Kansatz}) and \eqref{dg2KFinal} above. Furthermore, since we are interested in the amplitude mode in the $d$-wave channel in the expression for $\hat{g}_2^K$ we will consider $\delta\Delta_{\bn}^{L}(\omega)$ in the form
\beg\label{dDeltadwave}
\delta\Delta_\bn^L(\omega)=\gamma_\bn\delta\Delta_{\omega}^L.
\en

It is instructive to discuss two distinct contributions to \eqref{SelfEq}.
We start by considering the contribution to \eqref{Self} from $\delta\hat{g}_2^K$. In view of the comments above for the trace we find
\beg\label{TrAnom}
\begin{split}
\textrm{Tr}\left\{(-i\hat{\tau}_2)\delta\hat{g}_2^K(\bn\eps;\bq\omega)\right\}&=\frac{{\cal A}_\bn^K(\eps_+,\eps_-)\gamma_\bn\delta\Delta_{\bq\omega}^L}{\eta_{\bn\eps+\omega}^{R}+\eta_{\bn\eps-\omega}^{A}}\\&\times(t_{\eps+\omega}-t_{\eps-\omega}),
\end{split}
\en
where $\eps_{\pm}=\eps\pm\omega$ and 
\beg\label{calAK}
{\cal A}_\bn^K(\eps_{+},\eps_{-})=g_{\bn\eps_+}^Rg_{\bn\eps_-}^A+f_{\bn\eps_+}^Rf_{\bn\eps_-}^A+1.
\en
This contribution to the SH susceptibility originates from the non-equilibrium single particle distribution. 

It remains to evaluate the contribution from the first part of \eqref{g2Kansatz}, which obviously contains the equilibrium distribution function since $t_\eps=1-2n_F(\eps)$ and $n_F(\eps)$ is the Fermi-Dirac distribution function. Using \eqref{g2RAFinal} we have:
\beg\label{chinorm}
\begin{split}
&\textrm{Tr}\left\{(-i\hat{\tau}_2)\left[\delta\hat{g}_2^R(\bn\eps;0,\omega)t_{\eps-\omega}-\delta\hat{g}_2^A(\bn\eps;0,\omega)t_{\eps+\omega}\right]\right\}\\&=\left[\frac{{\cal A}_\bn^R(\eps_+,\eps_-)t_{\eps-\omega}}{\eta_{\bn\eps+\omega}^{R}+\eta_{\bn\eps-\omega}^{R}}-
\frac{{\cal A}_\bn^A(\eps_+,\eps_-)t_{\eps+\omega}}{\eta_{\bn\eps+\omega}^{A}+\eta_{\bn\eps-\omega}^{A}}\right]{\cal Y}(\bn)\delta\Delta_{\omega}^L, 
\end{split}
\en
where we introduced functions
\beg\label{ARA}
{\cal A}_\bn^{R(A)}(\eps_{+},\eps_{-})=g_{\bn\eps+\omega}^{R(A)}g_{\bn\eps-\omega}^{R(A)}+f_{\bn\eps+\omega}^{R(A)}f_{\bn\eps-\omega}^{R(A)}+1.
\en
It has to be noted that these expressions are fully analogous to those found for the $s$-wave case.\cite{Moore2017,Eremin2024,Yantao2024,Kamenev2025}
\subsection{Amplitude mode susceptibility}
Inserting the expressions above into the self-consistency equation \eqref{dDeltadwave} yields the linear consistency relation in the form $\chi_{\textrm{SH}}^{-1}(\Omega)\delta\Delta_{\Omega}^L=0$ where $\chi_{\textrm{SH}}^{-1}(\Omega)$ has a physical meaning of the inverse
longitudinal susceptibility and it is given by
\begin{widetext}
\beg\label{chiSHq0}
\begin{split}
\chi_{\textrm{SH}}^{-1}(\Omega)=-\frac{1}{\lambda}+\int\limits_{0}^{2\pi}\gamma_\bn^2\frac{d\phi_\bn}{2\pi}\int\limits_{-\omega_D}^{\omega_D}d\eps\left\{
\frac{{\cal A}_\bn^K(\eps_+,\eps_-)(t_{\eps+\Omega/2}-t_{\eps-\Omega/2})}{\eta_{\bn\eps+\Omega/2}^{R}+\eta_{\bn\eps-\Omega/2}^{A}}+\frac{{\cal A}_\bn^R(\eps_+,\eps_-)t_{\eps-\Omega/2}}{\eta_{\bn\eps+\Omega/2}^{R}+\eta_{\bn\eps-\Omega/2}^{R}}-
\frac{{\cal A}_\bn^A(\eps_+,\eps_-)t_{\eps+\Omega/2}}{\eta_{\bn\eps+\Omega/2}^{A}+\eta_{\bn\eps-\Omega/2}^{A}}\right\}.
\end{split}
\en
\end{widetext}
Here $\eps_{\pm}=\eps\pm\Omega/2$, $\Omega=2\omega$ and the dimensionless coupling constant is given by
\beg\label{Coupling}
\frac{1}{\lambda}=\frac{1}{\Delta}\int\limits_{-\omega_D}^{\omega_D}d\eps\int\limits_0^{2\pi}\gamma_\bn\frac{d\phi_\bn}{2\pi}\left(f_{\bn\eps}^R-f_{\bn\eps}^A\right)t_\eps,
\en
where $\Delta$ is the value of the order parameter in equilibrium. 
It is worth noting that after setting ${\cal Y}(\bn)=1$ we recover previously derived expression for the SH susceptibility for the $s$-wave superconductor.\cite{Phan2023,Pasha2025,Kamenev2025} Note also that while the integrals in Eqs. \eqref{chiSHq0} and \eqref{Coupling} need to be cut off at a Debye frequency $\omega_D$, being taken together they yield
a UV convergent integral. Thus expression for the inverse
susceptibility, $\chi_{\textrm{SH}}^{-1}(\Omega)$, is, in fact, cutoff independent. 

In Fig. 1 we show the plots of the real and imaginary parts of the function $\chi_{\textrm{SH}}(\Omega)$ for both $s$-wave and $d$-wave cases. 
\begin{figure}
\includegraphics[width=0.850\linewidth]{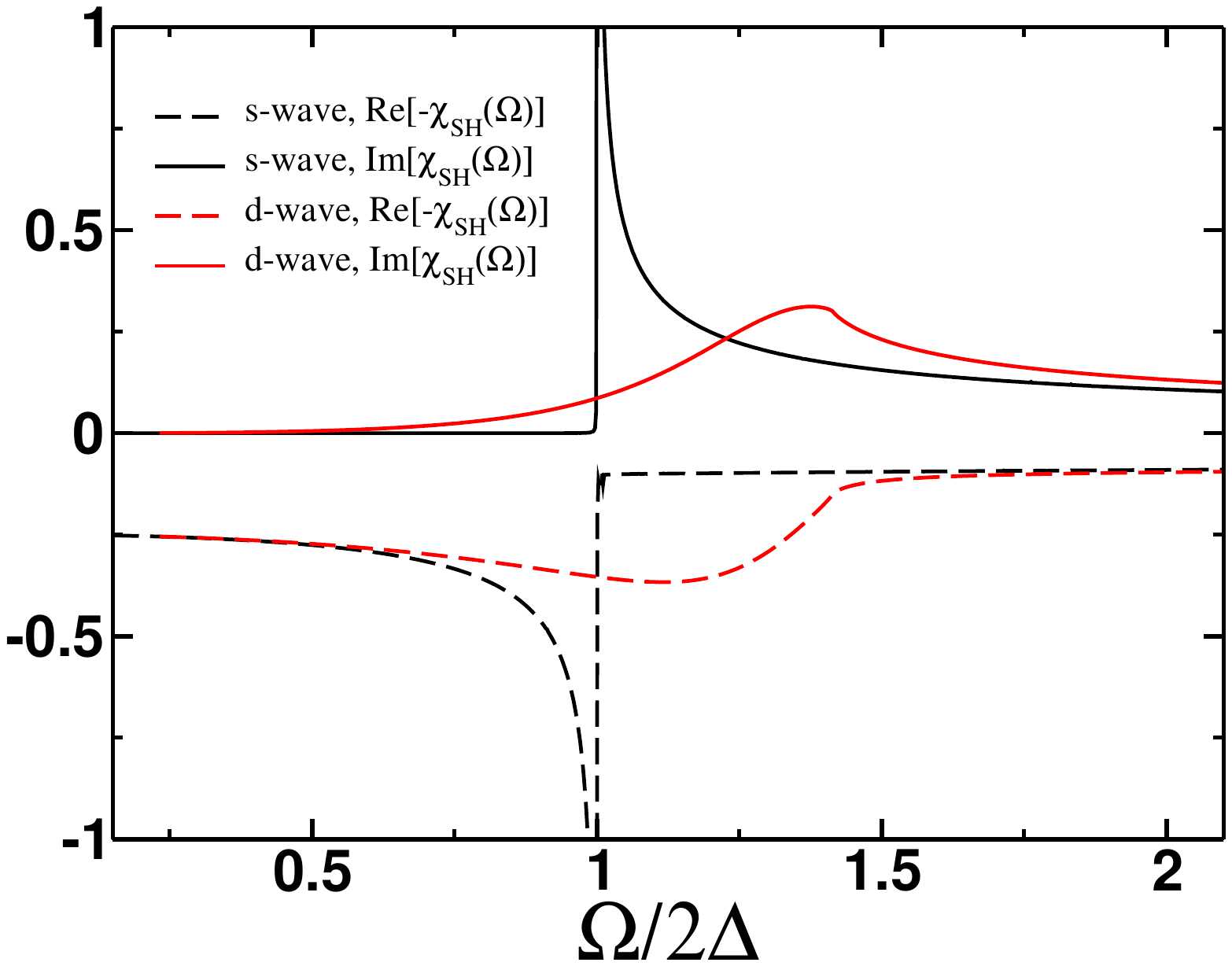}
\caption{Comparison between the real and imaginary parts of the function $\chi_{\textrm{SH}}(\Omega)$ for the $s$-wave and $d$-wave cases. For the $s$-wave case function $\mathrm{Im}[\chi_{\textrm{SH}}(\Omega)]$ exhibits a well-known Schmid-Higgs resonance at $\Omega=2\Delta$. In contrast for the $d$-wave pairing Schmid-Higgs susceptibility exhibits a maximum at frequency $\approx 2\sqrt{2}\Delta$.} 
\label{Fig1}
\end{figure}
The Schmid-Higgs resonance at $\Omega=2\Delta$ for the $s$-wave case is completely smeared out in the $d$-wave case. For this reason we already may expect that the dynamics of the SH mode will decay faster in the $d$-wave superconductor. Notably, the peak in the imaginary part of 
$\chi_{\textrm{SH}}(\Omega)$ is at the frequency $\omega_{\mathrm{peak}}>2\Delta$.
\begin{figure}
\includegraphics[width=0.90\linewidth]{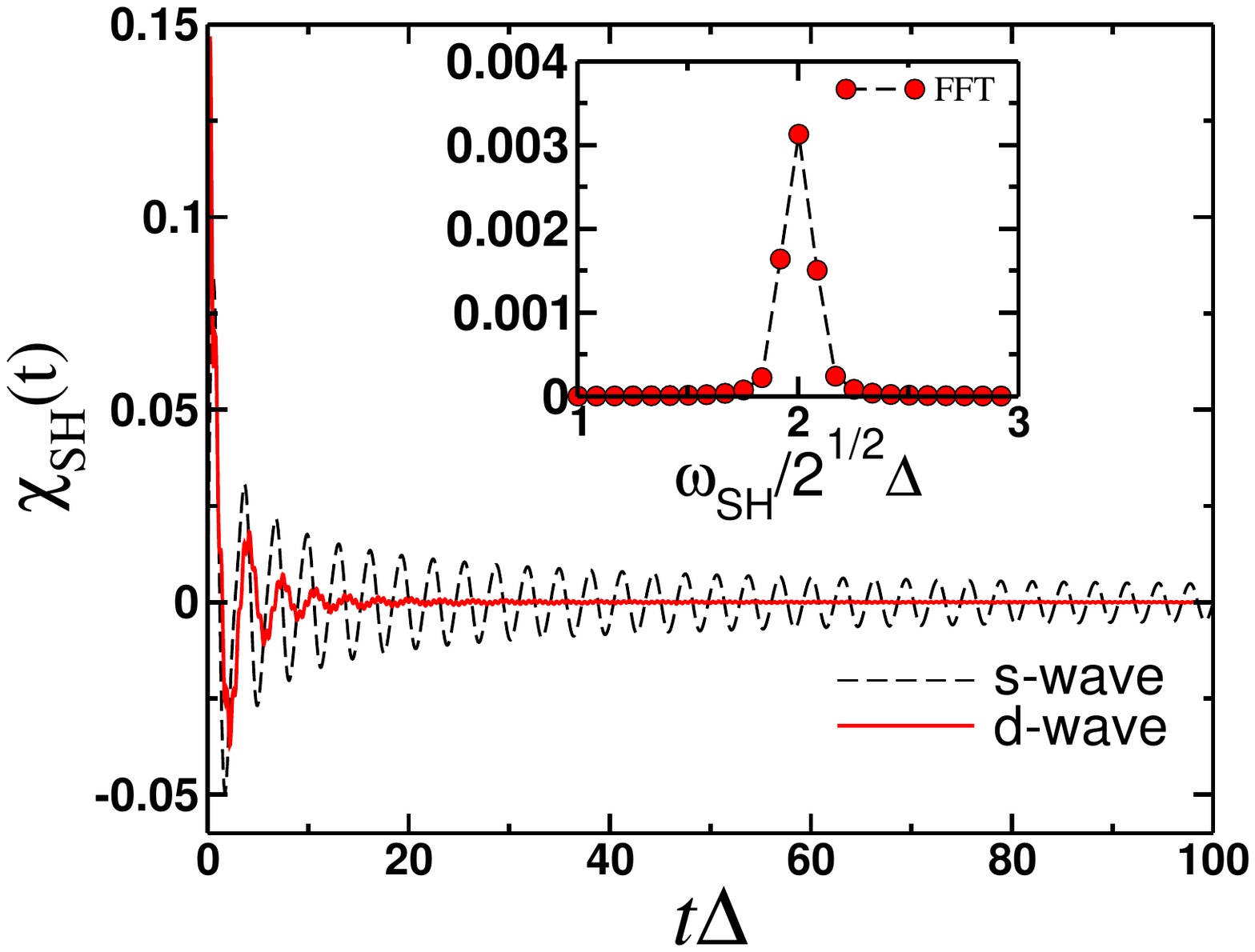}
\caption{Time dependence of the Schmid-Higgs susceptibility computed from the Fourier transform of the function $\chi_{\mathrm{SH}}(\Omega)$ for the $s$-wave and $d$-wave cases. The dynamics in the $d$-wave case decays much faster. We found that at long times in the $d$-wave superconductor $\chi_{\mathrm{SH}}(t)\sim 1/t^2$. Inset: the plot of the results of the Fast Fourier Transform (FFT) showing that the $d$-wave pair susceptibility oscillates with the frequency $2\sqrt{2}\Delta$. } 
\label{Fig2-CHIt}
\end{figure}

In Fig. \ref{Fig2-CHIt} we show the time dependence of the Schmid-Higgs susceptibility for both $s$- and $d$-wave superconductors which we have evaluated numerically by performing the Fourier transform of the function $\chi_{\mathrm{SH}}(\Omega)$. 
We find that indeed the SH susceptibility decays much faster in the $d$-wave case compared to the $s$-wave one. From our numerical analysis 
it follows that in the $d$-wave case is $\chi_{\mathrm{SH}}(t\Delta\gg1)\sim 1/t^2$. Furthermore, the frequency of the oscillations $\omega_{\mathrm{SH}}\approx 2\Delta_{\mathrm{an}}$, where $\Delta_{\mathrm{an}}=\sqrt{2}\Delta$ is the paring amplitude in the anti-nodal direction.

\section{Equations of motion for the Anderson pseudospins}
As it is well known, dynamics of the Schmid-Higgs mode can be directly deduced from the equations of motion for the Anderson pseudospins.
\cite{Anderson1958} Indeed since the second term in the Hamiltonian, Eq. \eqref{Eq1}, it can be conveniently re-written in terms of the operators 
\beg\label{Pseudospins}
\begin{split}
&\hat{S}_\bk^+=\hat{c}_{\bk\up}\dg\hat{c}_{-\bk\dn}\dg, ~\hat{S}_\bk^{-}=\hat{c}_{-\bk\dn}\hat{c}_{\bk\up},\\ 
&\hat{S}_\bk^z=\frac{1}{2}\left(\hat{c}_{\bk\up}\dg\hat{c}_{\bk\up}+\hat{c}_{-\bk\dn}\dg\hat{c}_{-\bk\dn}-1\right).
\end{split}
\en
These are familiar Anderson pseudospin operators \cite{Anderson1958} which satisfy the angular momentum commutation relations $[S^a_\bk,S^b_\bq]=i \epsilon^{abc}\delta_{\bk\bq}S^c_\bk$. Thus our initial model Hamiltonian (\ref{Eq1}) can be re-written as follows:
\beg\label{ASH}
\hat{H}=2\sum\limits_{\bk}(\veps_{\bk}-\mu)\hat{S}_\bk^z-\lambda\sum\limits_{\bk,\bq}\gamma_{\bk}\gamma_{\bq} \hat{S}_\bk^{+} 
\hat{S}_\bq^{-}.
\en
To obtain the ground state in the mean-field approximation, the pseudospin operators are replaced with their expectation values $\hat{S}_\bk^a\to\langle \hat{S}_\bk^a\rangle={S}_\bk^a$. As a result, the spin Hamiltonian (\ref{ASH}) 
becomes a classical Hamiltonian of the form
\beg\label{Hmf}
H_{\mathrm{cl.}}=\sum\limits_{\bk}{\vec B}_\bk\cdot{\vec S}_\bk, ~{\vec B}_\bk=2(-\gamma_\bk \Delta_x,-\gamma_\bk\Delta_y,\xi_\bk)
\en
where $\xi_\bk=\veps_\bk-\mu$ and $\Delta_{x,y}$ are the components of the complex pairing field 
\beg\label{Self}
\Delta^+(t)=\Delta_x(t)+i\Delta_y(t)=\lambda\sum_\bk\gamma_\bk S_{\bk}^+(t)
\en
and $S_{\bk}^{\pm}=S_\bk^x\pm iS_\bk^y$.
The time evolution of the pseudospin components along with the pairing field $\Delta^+$ is governed by the classical equations of motion which are obtained by evaluating the Poisson brackets of $S_\bk^a$ with the Hamiltonian (\ref{Hmf}):
\beg\label{EqMot}
\frac{\partial{\vec{S}}_\bk}{\partial t}=\vec{B}_\bk(t) \times \vec{S}_\bk(t).
\en
In the ground state, each pseudospin is aligned so that time derivatives in Eq. \eqref{EqMot} are identically zero. For simplicity we assume that in the ground state $\Delta=\Delta_x$, it follows
\beg\label{Equi}
S_\bk^x=\frac{\gamma_{\bk}\Delta }{2\sqrt{\xi_{\bk}^2 + |\gamma_{\bk}\Delta|^2}}, ~
S_\bk^z=-\frac{\xi_{\bk}}{2\sqrt{\xi_{\bk}^2 + |\gamma_{\bk}\Delta|^2}}
\en
and $S_\bk^y=0$. In addition to the self-consistency equation(s) for the pairing field, we also need to consider the particle number equation which fixes the value of the chemical potential:
\beg\label{ParticleNumber}
n=\sum_{\bk}\left(1-\frac{\xi_{\bk}}{\sqrt{\xi_{\bk}^2 + |\gamma_{\bk}\Delta|^2}} \right).
\en
The dynamics of the pairing field is induced by a sudden change of the pairing strength $\lambda\to\lambda'$.
As we have already mentioned above, for the $d$-wave superconductor equations (\ref{Equi}) have been analyzed numerically for an arbitrary magnitude of the quench.\cite{PRL2015} For the purposes of comparing the results of the previous calculation with the dynamics of the order parameter governed by (\ref{Equi}) we will limit the discussion to the so-called linear regime defined by $\delta_\lambda=|\lambda'-\lambda|\ll\lambda$.\cite{VolkovKogan1973,Yuzbashyan2015}

\begin{figure}
\includegraphics[width=0.850\linewidth]{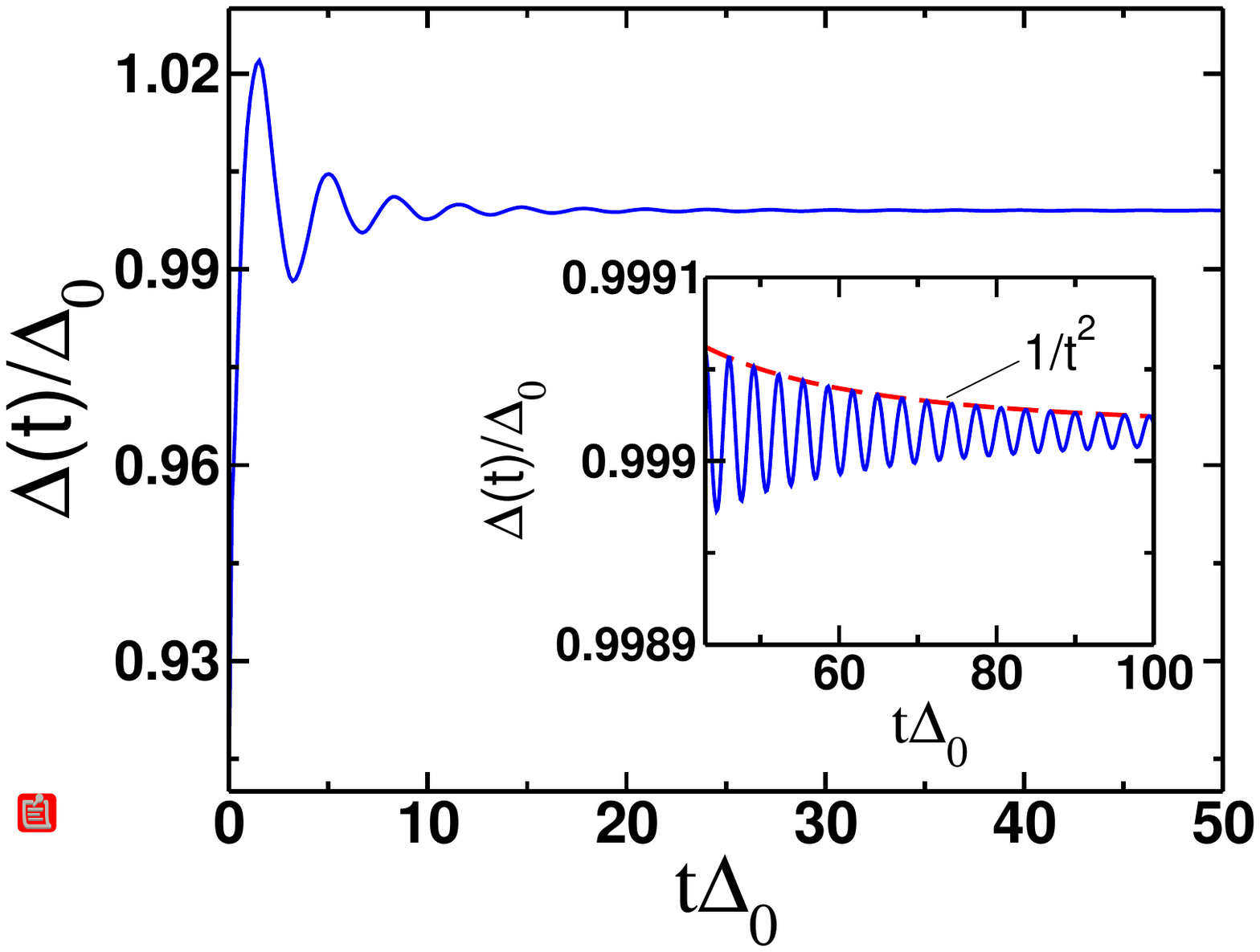}
\caption{Main panel: time dependence of the pairing amplitude following a small change of the pairing strength plotted in the units of equilibrium pairing amplitude $\Delta_0$. Inset: the fit of the $\Delta(t)$ at long times $t\Delta_0\gg 1$ showing that the oscillation amplitude decays as $\approx 1/t^2$. } 
\label{Fig3-DLTt}
\end{figure}

In Fig. \ref{Fig3-DLTt} we show an example of the order parameter dynamics initiated by a weak quench $\delta\lambda/\lambda=0.05$. By fitting the dependence of $\Delta(t)$ at long times we found 
\beg\label{dDLTt}
\Delta(t\gg\Delta_0^{-1})\approx\Delta_0\left[1+A\frac{\cos(2\omega_{\mathrm{SH}} t+\pi/4)}{(t\Delta_0)^2}\right]
\en
with $A\ll 1$ and
\beg\label{AntiNodal}
\omega_{\mathrm{SH}}=\sqrt{2}\Delta_0\equiv\Delta_{\mathrm{an}}.
\en
Thus we confirm that the frequency of the Schmid-Higgs mode in the case of the $d$-wave superconductor is determined by the value of the order parameter $\Delta_{\mathrm{an}}$ in the anti-nodal direction. It is important to note here that our result \eqref{AntiNodal} differs from those reported in Refs. \cite{BarlasVarma2013,Wu2020}. We believe that the difference stems from the fact that these works used the unnormalized $d$-wave form factor $\cos 2\phi$ rather than the normalized $\sqrt{2}\cos 2\phi$ one  (see Eq. \eqref{NormGamma}), which likely lead to a quantitative shift in the resulting mode energy due to modified angular averages entering the response kernel.

\section{Discussion and Conclusions}
It is well known that even small amounts of potential impurities are detrimental to unconventional superconductivity.\cite{MineevSamokhin} Indeed, for the case of uncorrelated point-like potential impurities, instead of expressions (\ref{etaeps}) for the correlation functions one will find similar expressions in which one will have to replace $\eps\to\eps+ig/2\tau$ and $\Delta_\bn\to\Delta_\bn-if/2\tau$ where $\tau^{-1}$ is the disorder scattering rate. After inserting these expressions into self-consistency condition, one discovers that pairing amplitude  
is quickly suppressed with an increase in $\tau^{-1}$.
For this reason in all the calculations above it was assumed that a system is sufficiently clean. An example of such a system would be CeCoIn$_5$ or CeCu$_2$Si$_2$.\cite{Petrovic2001,Movshovich2001,Sarrao2007,Steglich1979,Miyake2007,Morr2014}  In addition, disorder effects will inevitably change the decay rate of the collective mode oscillations. It is likely that the effect of pair breaking on the time-dependent pair susceptibility at $\bq=0$  will be similar to the effect of keeping finite value of $\bq$ but in a clean superconductor just like it happens in the $s$-wave case.\cite{Kamenev2025}

In our study we kept only diagonal in momentum terms in the pairing Hamiltonian which is justified as long as we are interested in the dynamics at $\bq=0$.
The question of spatially resolved dynamics of the Schmid-Higgs mode in $d$-wave superconductors is an interesting one. It is by now well-known \cite{Pasha2025,Kamenev2025} that in the $s$-wave case the Schmid-Higgs mode decays at $1/t^2$ at finite momentum which is in contrast with $\sim 1/\sqrt{t}$ behavior when $\bq=0$. With the $d$-wave superconductor there are two distinct directions of $\bq$ - nodal and anti-nodal one - and therefore one may generally expect that the time-dependence of the pair susceptibility will be different for these two cases.\cite{Unpublished}

In this work we have considered the short time dynamics of the collective pairing mode (Schmid-Higgs mode) in an unconventional superconductor with $d$-wave symmetry. Specifically, assuming that the dynamics has been initiated by small deviations from equilibrium, we computed the pairing susceptibility and also solved for the dynamics of the order parameter directly using the equations of motion within the mean-field theory approach. We have found that the frequency of the Schmid-Higgs mode is determined by the value of the pairing gap in the anti-nodal direction. In addition, we found that the amplitude of the collective mode oscillations decay as a power law according to $1/t^2$.

\section{Acknowledgments}
We would like to thank Kazi Ranjibul Islam and Andrey Chubukov for stimulating discussions. Our work was financially supported by the National Science foundation grant DMR-2400484. One of us (MD) has performed the main part of this  work at Aspen Center for Physics, which is supported by the National Science Foundation grant PHY-2210452. 


\end{document}